**Contributions of Albert Einstein to Earth Sciences: A review in Commemoration of the World Year of Physics**

Jesús Martínez-Frías[1], David Hochberg[1] and Fernando Rull[1,2]

[1] *Centro de Astrobiologia (CSIC/INTA), associated to the NASA Astrobiology Institute, Ctra de Ajalvir, km 4, 28850 Torrejon de Ardoz, Madrid, Spain, Tel: +34-91-5201111, Fax: +34-91-5201621, e-mail: martinezfrias@mncn.csic.es*

[2] *Cristalografía y Mineralogía, Facultad de Ciencias and Unidad Asociada al Centro de Astrobiologia, Universidad de Valladolid–CSIC, 47006 Valladolid, Spain*

**Abstract**

The World Year of Physics (2005) is an international celebration to commemorate the 100th anniversary of Einstein's "*Annus Mirabilis*". The United Nations has officially declared 2005 the International Year of Physics. However, the impact of Einstein´s ideas was not restricted to physics. Among numerous other disciplines, Einstein also made significant and specific contributions to Earth Sciences. His geosciences-related letters, comments, and scientific articles, are dispersed, not easily accessible and are poorly known. The present review attempts to integrate them, as a tribute to Einstein in commemoration of this centenary. These contributions can be classified into three basic areas: geodynamics, geological (planetary) catastrophism and fluvial geomorphology. Regarding geodynamics, Einstein essentially supported Hapgood's very controversial theory called Earth Crust Displacement. With respect to geological (planetary) catastrophism, it is shown how the ideas of Einstein about Velikovsky's proposals evolved from 1946 to 1955. Finally, in relation with fluvial geodynamics, the review incorporates the elegant work in which Einstein explains the formation of meandering rivers. A general analysis of his contributions is also carried out from today's perspective. Given the interdisciplinarity and implications of Einstein's achievements to multiple fields of knowledge, we propose that the year 2005 serves, rather than to confine his universal figure within a specific scientific area, to broaden it? for a better appreciation of this brilliant scientist in all of his dimensions.

**1   Introduction**

3         In 1905, the same year that Albert Einstein obtained his doctorate after
4   submitting his thesis (University of Zurich) "*On a new* determination of molecular
5   dimensions", he published his five famous articles (Pais, 1983; Stachel, 1998; Bushev,
6   2000): *On an heuristic viewpoint about the emergence and conversion of light* -
7   submitted in March; *A new determination of the molecular dimensions* – submitted in
8   April; *On the movement, required by the molecular theory of heat, of particles*
9   *suspended in a motionless fluid* - submitted in May; *Towards the electrodynamics of*
10  *moving bodies* - submitted in June, and *Does the inertia of a body depend on the energy*
11  *it contains ?* - submitted in September. The World Year of Physics (2005) is an
12  international celebration to commemorate the $100^{th}$ anniversary of Albert Einstein's
13  "*Annus Mirabilis*". It also marks the $50^{th}$ anniversary of Einstein's death (McCrea and
14  Lawson, 1955). The United Nations has officially declared 2005 the International Year
15  of Physics (WYP, 2005; APS, 2005; Sathyasheelappa, 2005).

17        However, the impact of Einstein's theories was not restricted to physics and
18  technological applications of physics: Einstein's achievements influenced philosophy,
19  art, history of science, literature, and many other disciplines (Hentschel, 1986; Byrne,
20  1980; Goenner and Castagnetti, 1996). Beyond the general implications for geophysics
21  and other geosciences of his revolutionary ideas on magnetism and gravity, Einstein
22  also made significant, and much more specific, contributions to Earth Sciences. Even
23  one could speculate that such interest, particularly related to some geomorphological
24  fluvial problems, could mark, in a certain way, his son's vocation: a renowned hydraulic engineer, who is famous in his own right for his sediment transport equation (Ettema and Mutel 2004).

However, these works are dispersed and poorly known, essentially because: a) they were eclipsed by his major scientific accomplishments, and b) his geosciences-related notes, letters, comments, and even scientific articles are disperse (not constituting nor collected together as a thematic whole). In fact, to the best of our knowledge, and after conducting an exhaustive search in the ISI-Web of Science, the present work is the first scientific review article which brings together all this geological information. These contributions can be classified into three basic areas: geodynamics, geological (planetary) catastrophism and fluvial geomorphology.

**Geodynamics**

Geodynamics is the study of the processes and mechanisms that shape, and have shaped, our planet in the past, from the longest to the shortest length- and time-scales: from plate motions to microscopic crystal deformation and magma migration, and from millions of years to minutes.

Einstein's scientific considerations on this topic are principally connected to crustal geodynamics, and expounded in a series of correspondences in the mid 1950's with Charles Hapgood (*Courtesy of the Einstein Archives Online - http://www.alberteinstein.info/*). Hapgood (1958) developed a controversial theory called Earth Crust Displacement (ECD), according to which the earth's lithosphere can sometimes move as a whole over the asthenosphere.

The proposed mechanism for these crustal displacements is related to the build-up of thick ice-sheets in polar and near-polar regions. This idea is a global geological theory which, in conjunction with many other anomalies of earth science, and in a sophisticated way (but also controversially) (Martinez-Frias and Barrera, 2000) attempts to elucidate how and why large parts of Antarctica could have remained ice-free until recently. Hapgood claimed that towards the end of the last ice age, the widespread mass of glacial ice covering the northern continents provoked the lithosphere to 'slip' over the asthenosphere, moving Antarctica, for several centuries, from a location in the middle latitudes to its current position and, in unison, rotating the other continents. Antarctica's movement to the polar region precipitated the growth of its ice cap. Similarly, by shifting the northern ice sheets out of the arctic zone, the end of the ice age was made possible. Support for this theory was given by Einstein,

*"I frequently receive communications from people who wish to consult me concerning their unpublished ideas. It goes without saying that these ideas are very seldom possessed of scientific validity. The very first communication, however, that I received from Mr Hapgood electrified me. His idea is original, of great simplicity, and - if it continues to prove itself - of great importance to everything that is related to the history of the earth's surface."* (Einstein, 18th of May 1954, courtesy of the Einstein Archives Online).

The following Einstein quotation illustrates very well how Einstein was knowledgeable and aware of the geodynamical intricacies:

*"In a polar region there is a continual deposition of ice, which is not symmetrically distributed about the pole. The earth's rotation acts on these unsymmetrically deposited masses, and produces centrifugal momentum that is transmitted to the rigid crust of the earth. The constantly increasing centrifugal momentum produced in this way will, when it has reached a certain point, produce a movement of the earth's crust over the rest of the earth's body, and will displace the polar regions toward the equator."* (Hapgood 1958, p. 1)

It is important to note that, although Einstein wrote this, he had doubts that the weight of the ice-caps would have been sufficient to move the crust. Thus, in endorsing Hapgood's work, Einstein however, did so with some lingering concerns.

*"Without a doubt the earth's crust is strong enough not to give way proportionately as the ice is deposited. The only doubtful assumption is that the earth's crust can be moved easily enough over the inner layers".*

Nevertheless, Albert Einstein wrote about Charles Hapgood's theory of Earth Crust Displacement in a letter to William Farrington (Flem-Ath, 2002) of the Department of Geology and Mineralogy at the University of Massachusetts: "*I think that the idea of Mr. Hapgood has to be taken quite seriously.*"

Hapgood and Einstein continued to correspond and finally met in January of 1955. Einstein's last letter to Hapgood was dated the 9th of March 1955 just weeks before he died on the 18th of April 1955.

What is the situation concerning these ideas from today's perspective? It appears to be clear that the geologic revolution that took place in the 1960s -explicitly the development of plate tectonics- overshadowed Hapgood's theory. The data demonstrating that the poles were in different positions (very slight shifts) during the Pleistocene is convincing (Eden 2005), and explained why Hapgood's theory was seriously considered by scientists such as Einstein and K.F. Mather (a prestigious geomorphologist and paleontologist). But plate tectonics had unified many of the features and characteristics of seafloor spreading and continental drift and into a coherent model and had 'renovated' geologists' understanding of ocean basins, continents, mountains, and earth history. In fact, Hapgood himself first realized the huge amount of evidence regarding the high speed at which the shift of the poles appeared to have occurred; speed that a sole mechanism of continental drift was unable to explain. Thus, Hapgood saw himself obliged to adhere to a much slower shift of the poles, accepting plate tectonics to uphold the support of the scientific (mainly geological) community. For instance, a significant geological problem comes from the classical concept of "isostasy": "the balance or equilibrium between adjacent blocks of crust resting on a plastic mantle" (Plummer and McGeary, 1996). Isostatic rebound affects the rise or fall of sea levels (e.g. Krause, 1996) and Earth Crust Displacement provides no satisfactory explanation to this problem. In fact, this same author considers that Einstein's claim in Hapgood (1958) that, at a certain critical point, a slip of the earth's crust is bound to occur due to an irregularly distributed icemass that "fails" to take isostasy into consideration. Following Krause (1996): a) between isostasy and the tendency of ice to flow plastically, the critical point mentioned by Einstein is never reached, and b) the earth's crust is not rigid, as Einstein stated. Instead, as ice builds up on a landmass, that landmass is depressed an appropriate amount to carry the load. Also, assuming ECD could take place, it seems reasonable that close to the poles of rotation there should have been some form of augmented geologic activity, such as faulting or volcanism, due to increased friction between the lithosphere and asthenosphere (Krause, 1996).

Recently, Brass (2002), alluding to recent results obtained by the CLIMAP Project (1981), claimed that "Earth's lithosphere is attached to the mantle in such a way as to make Earth Crustal Displacement impracticable" (Brass, 2002, p.45). Others have gone further, labeling the ECD hypothesis as pseudoscience (Earle, 2003), and emphasizing that it is mainly based on circumstantial evidence and that it reflects a clear misunderstanding of the numerous well-documented observations of the earth's crust and mantle with which it is inconsistent.

In conclusion, a general consensus appears to imply that ECD itself is, from a geodynamic point of view, not well thought-out, is unable to address central geologic questions, and is contrary to current geological knowledge.

**Geological (planetary) catastrophism**

Catastrophism is the theory that the Earth has been affected by rapid, short-lived, violent events that were sometimes worldwide in range. The theory was systematized and defended by Georges Cuvier (1769–1832), whose position, as one of the greatest geologists of his epoch, overrode all opposition. It was strongly contested by George Poulett Scrope (1797-1876) and, in particular, by Sir Charles Lyell (1797–1875), under whose scientific authority the contrary doctrine (uniformitarianism) gradually became more popular (and scientifically much more accepted). Recent theories of meteorite, asteroid, or comet impacts triggering mass extinctions can be interpreted as a certain revival of catastrophism (Rupke, 1992; Davies, 1993).

In 1940, Immanuel Velikovsky (1895-1979) studied a number of natural disasters, described in the Bible (e.g. the parting of the Red Sea). By comparison with similar events described in Egyptian texts, Velokovsky (1950) postulated that the events related to the same catastrophes. He concluded that many of these events were linked to a single global cataclysm, and that Venus was involved. Velikovsky's (1950) book

Worlds in Collision, claims, among many other things, that the planet Venus did not exist until recently.

He proposed that a huge internal convulsion of the planet may have caused Jupiter to outwardly disgorge (or disrupt the orbit of) a comet, which fell towards the sun in a highly elliptical orbit. It passed close to Mars, dragging it out of its orbit and pulling away its atmosphere, then went by the Earth, causing severe catastrophes, then returned again, after going around the sun in a 52 year cycle, provoking more havoc. It finally settled in a relatively stable, circular orbit, to become the planet Venus as we know it. Of course, Velikovsky was hostilely opposed by the vast majority of the scientific community, mainly because he did not provide scientific evidence for his claims. On the contrary, his ideas were mostly based on assuming that cosmological facts must conform to mythology. Velikovsky's theories were considered to be pseudoscience at best, and sheer nonsense at worst. In fact, Velikovsky was, for nearly ten years, persona no grata on college and university campuses (MacKie, 2005).

Curiously, Albert Einstein, after engaging in long discussions with Velikovsky (Kogan and Sharon, 1999) at least respected his effort and some of his views, even if he did not fully accept his proposals. He showed:

a) his marked skepticism, stressing the weakness of Velikovsky's arguments

*"…I have read the whole book about the planet Venus. There is much of interest in the book which proves that in fact catastrophes have taken place which must be attributed to extraterrestrial causes. However it is evident to every sensible physicist that these catastrophes can have nothing to do with the planet Venus and that also the direction of the inclination of the terrestrial axis towards the ecliptic could not have undergone a considerable change without the total destruction of the entire earth's crust. Your arguments in this regard are so weak as opposed to the mechanical-astronomical ones, that no expert will be able to take them seriously. It were best in my opinion if you would in this way revise your books, which contain truly valuable material. If you cannot decide on this, then what is valuable in your deliberations will*

*become ineffective, and it may be difficult finding a sensible publisher who would take the risk of such a heavy fiasco upon himself…"* (8 July 1946, Kogan and Sharon, 1999).

b) his unambiguous disagreement concerning the confusion of facts and assumptions

*"…The reason for the energetic rejection of the opinions presented by you lies not in the assumption that in the motion of the heavenly bodies only gravitation and inertia are the determining factors. The reason for the rejection lies rather in the fact that on the basis of this assumption it was possible to calculate the temporal changes of star locations in the planetary system with an unimaginably great precision. Against such precise knowledge, speculations of the kind as were advanced by you do not come into consideration by an expert…"* (27 August 1952, Kogan and Sharon, 1999).

c) certain anger against the behavior of some scientists who criticized Velikovsky's ideas using preconception and non-scientific arguments. Einstein also began to accept catastrophism to explain some concrete events

*"…He can easily come to the opinion that you yourself don't believe it, and that you want only to mislead the public. I myself had originally thought that it could be so. This can explain Shapley's [Harlow] behavior, but in no case excuse it. This is the intolerance and arrogance together with brutality which one often finds in successful people, but especially in successful Americans. The offence against truthfulness, to which you rightly called my attention, is generally human, and in my eyes, less important. One must however give him credit that in the political arena he conducted himself courageously and independently, and just about carried his hide to the marketplace. Therefore it is more or less justified if we spread the mantle of Jewish neighborly love over him, difficult as it may be. To the point, I can say in short: catastrophes yes, Venus no…"* (22 May 1954, Kogan and Sharon, 1999).

d) final remarks which even emphasized the talent of Velikovsky

*"…you have presented me once more with the fruits of an almost eruptive productivity. I look forward with pleasure to reading the historical book that does not bring into danger the toes of my guild. How it stands with the toes of the other faculty, I do not know as yet. I think of the touching prayer: "Holy St. Florian, spare my house, put fire to others!" "…I have already carefully read the first volume of the memoirs to "Worlds in Collision," and have supplied it with a few marginal notes in pencil that can*

*be easily erased. I admire your dramatic talent and also the art and straightforwardness of Thackerey who has compelled the roaring astronomical lion to pull in a little his royal tail without showing enough respect for the truth…"* (17 March 1955, Kogan and Sharon, 1999).

All this confirms that, during nearly nine years of correspondence, there is a clear evolution of Einstein's thoughts about Velikovsky. This, far from being unusual, denotes rather Einstein's greatness. As Millikan (1949) said about Einstein, "I came to admire him most for his extraordinary open-mindedness, his modesty, his honesty, and his complete readiness to admit that he had been wrong and to change his position entirely in the light of new conditions (p.345)"

Today, more than a half century after Velikovsky's "Worlds in Collisions", a detailed review of the publications regarding catastrophism in the ISI-Web of Science shows that there are more than 50 contributions on this topic over the last 30 years and 46 specific records about Velikovsky. There is today a new interest in catastrophism (neo-catastrophism), but for most geoscientists, it represents only a reappraisal of the significance of some geological processes in terms of frequency and magnitude (even unique events or singularities). In other aspects, neo-catastrophism is being used in a distorted way by creationists and curiously many neo-catastrophists use myth to inform our understanding of the ancient sky, but reject Velikovsky's colliding planets hypothesis. Many Velikovskians now see "Worlds in Collision" as seriously flawed, if not completely wrong (Ellenberger, 1986), and today awareness of Velikovskian studies is mainly to be found in four groups: (1) the so called Saturnists (James, 1999); (2) The Velikovskian (http://www.velikovskian.com/index.htm); (3) the Society for Interdisciplinary Studies in Great Britain (http://www.knowledge.co.uk/sis/), and (4)

The Velikovsky Archive: (http://www.varchive.org). Burns (2003) has probably made the best and most realistic compilation of the present criticisms of Velikovsky's ideas and catastrophism in general, including views by, among others, prestigious scientists such as David Morrison and Clark Chapman (Morrison & Chapman, 1989), among others. In any event, Einstein's "support" for Velikovsky was more a human reaction against the manoeuvres of some scientists to discredit him rather than an unambiguous scientific acceptance of his ideas. What Einstein really supported was Velikovsky's right as a scientist to question the established model, and to replace it if another model is demonstrated to be more correct.

**Fluvial Geomorphology**

Geomorphology is the study of the Earth's landscapes and landforms, the processes by which the landforms originated, their age, and the nature of the materials underlying them. Fluvial geomorphology is the study of landforms and processes associated with rivers. Einstein was the first to articulate how helical flow helps to determine meander length and to promote down-current migration of the meandering rivers (Einstein, 1926; Bowker, 1988)

According to Baer's law, rivers in the northern hemisphere will erode mostly on the right bank, due to the rotation of the planet, leading to steeper right banks. Therefore, over time, the river bed will move to the right. Likewise, secondary (helical) flows due to an imbalance of the centrifugal forces near the boundaries are a very general topic in fluid mechanics, known as Ekman layers (1905). Knowing both postulates, Einstein (1926) elegantly explained in this journal how helical flow develops in a meandering river, and that because the higher-velocity segments of the stream will be driven to the outside (concave) part of the river curve, erosion will be greater just there. He also noted that due to the inertia of the helical flow, the circulation, and related erosion, will attain a maximum beyond the inflection of the bend. Therefore, the wave-form of the river will migrate in a down-current direction. Finally, Einstein (1926) established that the larger the cross-sectional area of a river, the slower the helical flow will be absorbed by friction. Thus, larger rivers comprise meandering structures with longer wavelengths. As the scientist explains:

> *"…I begin with a little experiment which anybody can easily repeat. Imagine a flat-bottomed cup full of tea. At the bottom there are some tea leaves, which stay there because they are rather heavier than the liquid they have replaced. If the liquid is made to rotate by a spoon, the leaves will soon collect in the center of the bottom of the cup. The explanation of this phenomenon is as follows: the rotation of the liquid causes a centrifugal force to act on it. This in itself would give rise to no change in the flow of the liquid if the latter rotated like a solid body. But in the neighborhood of the walls of the cup the liquid is restrained by friction, so that the angular velocity with which it rotates is less there than in other places nearer the center. In particular, the angular velocity of rotation, and therefore the centrifugal force, will be smaller near the bottom than higher up. The result of this will be a circular movement [helical flow] of the liquid of the type illustrated in [the figure] which goes on increasing until, under the influence of ground friction, it becomes stationary. The tea leaves are swept into the center by the circular movement and act as proof of its existence…"* (Einstein, 1926, p.250)

From today's perspective, Einstein's third contribution is obviously different from the two others discussed here. It is not a comment, a letter, or the text fragment of a book preface. It is a bona-fide scientific article by Einstein. Hence its present unambiguousness and its geological and geomorphological utility.

There are actually two statements in Einstein's paper on meanders: (1) the Coriolis force is responsible for the onset of a meander (this is related to the Baer's law). (2)

once initiated, the secondary flow is responsible for the amplification of the meander, whatever its curvature (right or left). While (2) is certainly correct, (1) is almost not discussed in Einstein's paper, and is far from obvious. The influence of the Coriolis force is usually characterized by a number, the Rossby number, which compares the inertia to the Coriolis force. Except for large-scale flows (ocean, atmosphere), Earth rotation is often negligible. This number for a river is typically 10-1000 (it may be smaller near large estuaries). As a consequence, although Einstein's argument about the amplification and the migration of a meander is essentially correct, its relation with the Coriolis force is, in most cases, negligible.

However, although Einstein provided a physical explanation of flow around bends in sinuous channels that survives to this day, it has been largely overlooked. A perusal of texts and review papers on fluvial channels and their behavior failed to locate a single reference to Einstein's paper. John Allen fails to recognize Einstein's contribution. In his classic 1965 review paper on alluvial systems Allen, despite presenting a seemingly exhaustive review of the available literature, fails to cite Einstein (1926). In a subsequent reflective paper (Allen, 1978) he chooses to highlight the contribution of van Bendegom (1947) to meander studies, but again fails to cite Einstein (1926). Miall (1978) presents a historical review of fluvial sedimentology that also fails to cite Einstein's contribution. Likewise, examination of established and current texts, both sedimentological and hydraulic, again fails to produce a citation to Einstein (1926). The idea that bend migration is influenced, in a directional sense, by the Coriolis force has not found great support amongst subsequent fluvial geomorphologists, but it is widely used by researchers concerned with submarine, turbidity current channels to explain why the differential height of levees bounding the channels varies according to the hemisphere. Once again, papers making this observation and interpretation fail to cite Einstein's contribution.

To our knowledge, the most significant and recent references to his explanation regarding meandering rivers are those from Liverpool and Edwards (1995) and Moisy et al. (2001). The first presents a statistical model of an alluvial meandering river that is motivated by the physical nonlinear dynamics of river channel migration and by describing heterogeneity of the alluvial plain terrain by noise. The second uses Einstein's paper to illustrate the ubiquity of swirling flows at all scales in nature: spiral galaxies, atmospheric or oceanic circulation, bathtub vortices, or even stirring tea in a cup!

**Discussion/Conclusions**

Countless articles and volumes have been written about Einstein, and commemoration of the World Year of Physics is an excellent and appropriate initiative. Einstein is universally recognized as one of the greatest and most remarkable scientists of all times. It is a fact that Albert Einstein's contributions to science have opened up a new way of thinking about the universe, and the ramifications of his outstanding theories come into direct contact with many fields of knowledge; here, a review of his influence and contributions to earth sciences has been presented. But precisely for the interdisciplinarity of his achievements, it is recommended that 2005 serves, rather than to confine his universal figure within the boundaries of a specific scientific area, to broaden it for a better knowledge of the scientist and genius in all of his dimensions. As Einstein said (Perry, 2004):

"All religions, arts and sciences are branches of the same tree."

**Acknowledgements**

We acknowledge the courtesy of the Einstein Archives Online. We thank three anonymous referees, whose remarks and comments have contributed to greatly improve the original manuscript, and the fourth referee, Professor Elliot, for his extremely useful scientific remarks regarding the helical flow in sinuous rivers and its consequences which were incorporated verbatim into the article. We specially acknowledge the constant help and support received from Professor Juan Perez Mercader.